\documentclass[]{cupconf}
\usepackage{graphicx}

\title[Ellipticals and Bulges]{Chemical Evolution Models of Ellipticals and 
Bulges}

\author[Francesca Matteucci]{\\Francesca Matteucci$^{1}$}
\affiliation{$^1$
Department of Astronomy, University of Trieste\\
Via G.B. Tiepolo 11\\
34100 Trieste, Italy}

\begin{document}
\maketitle

\begin{abstract}
We review some of the models of chemical evolution of ellipticals and bulges of spirals. In particular, we focuse on the star formationn histories of ellipticals and their influence on chemical properties such as the [$\alpha$/Fe] versus [Fe/H], galactic mass and visual magnitudes. By comparing models with observational properties, we can constrain the timescales for the formation of these galaxies. The observational properties of stellar populations suggest that the more massive ellipticals formed on a shorter timescale than less massive ones, in the sense that both the star formation rate and the mass assembly rate, strictly linked properties, were more efficient in the most massive objects. Observational properties of true bulges seem to suggest that they are very similar to ellipticals and that they formed on a very short timescale: for the bulge of the Milky Way we suggest a timescale of 0.1 Gyr. This leads us to conclude that the Bulge evolved in a quite independent way from the galactic Disk.

\end{abstract}

\section{Introduction}
Galaxy formation is still an open subject and different scenarios have 
been proposed for the formation of elliptical galaxies.
Originally, Toomre \& Toomre (1972) suggested that ellipticals 
can form  from major mergers of massive disk galaxies.
Later on, Larson (1974), followed by many others 
(Arimoto \&  Yoshii 1987; 
Matteucci \& Tornamb\`e 1987; Bressan et al. 1994; Gibson, 1995; 
Pipino \& Matteucci, 2004; Merlin \& Chiosi, 2006) suggested an early 
monolithic collapse  of a gas cloud or early merging
of lumps of gas where dissipation plays a fundamental role and star formation 
stops after the occurrence of a galactic wind, as the main mechanism for the 
formation of ellipticals. In the following years Bender et al. (1993) suggested 
early merging of lumps containing gas and stars in which some 
energy dissipation is present. In more recent times a great deal of interest 
was devoted to the bottom-up scenario for galaxy formation,
expected from the cold dark matter (CDM) model, in which ellipticals 
should form by merging of 
early formed stellar systems in a wide redshift range and preferentially at 
late epochs with also some star formation 
associated with the merger events (e.g. Kauffmann et al. 1993). 
More recently, Bell et al. (2004) invoked 
dry merging at low redshift for the formation of ellipticals, namely merging 
of quiescent objects with no associated star formation. Bulges of spirals in most cases seem to follow the same 
properties as ellipticals (see Jablonka, this conference).
In the following we will try to see whether any of these proposed scenarios 
can account  for the majority of the observational constraints of ellipticals and bulges.

\section{Observational properties of elliptical galaxies}

From the observational point of view,  elliptical galaxies are characterized 
by spanning a large range in luminosities and masses, by containing mainly 
red giant stars and no gas (for a recent exhaustive review on elliptical 
galaxies we address the reader to Renzini, 2006).

In addition:
\begin{itemize}
\item
ellipticals are metal rich galaxies with mean stellar metallicity in the range
$<[Fe/H]>_{*}$= (-0.8-+0.3) dex (Kobayashi \& Arimoto 1999) and they are characterized 
by having large [$\alpha$/Fe] ratios 
($<[Mg/Fe]>_{*} > 0$,  from 0.05 to + 0.3 dex) 
in nuclei of giant ellipticals (Peletier 1989;
Worthey et al. 1992; Weiss et al. 1995; Greggio 1997; 
Kuntschner et al.  2001; Trager, 
this conference). This fact indicates that these galaxies and especially 
the most massive ones had a
short duration of formation ($\sim$ 0.3-0.5 Gyr, Matteucci 1994; 
Weiss et al. 1995): in fact, in order to have high $<[Mg/Fe]>_{*}$ 
ratios in their dominant stellar population, the supernovae (SNe) type Ia, 
which occur on a large interval of timescales, should not have had time to 
pollute significantly the interstellar medium (ISM) before the end of 
the star formation.   

\item Abundance gradients exist in ellipticals with typical metallicity 
gradients of ${\Delta [Fe/H] \over \Delta log r} =-0.3$ 
(see Kobayashi \& Arimoto, 1999, for a compilation of gradients and 
references therein).
These abundance gradients in the stellar populations are well reproduced 
by ``outside-in'' models for the formation of ellipticals as suggested by 
Martinelli et al. (1998), Pipino \& Matteucci (2004, hereafter PM04) and 
Pipino et al.(2006). It is not clear whether there is a 
correlation between abundance gradients and galactic mass 
(see Ogando et al. 2005, for a recent paper), as required by the classic 
monolithic model of Larson.

\item The tightness of the color-central velocity dispersion 
relation found for Virgo and Coma ellipticals
(Bower et al. 1992) also pointed to a short process of galaxy formation 
($\sim$ 1-2 Gyr). Bernardi et al. (1998) extended this conclusion also to field
ellpticals, but derived a timescale of galaxy formation slightly longer 
($\sim 2-3$ Gyr).

\item The thinness of the fundamental plane seen edge-on (M/L versus M)
for ellipticals in the
same two clusters (Renzini \& Ciotti 1993) indicates again a short process 
for the formation of stars in these galaxies.

\item Another very interesting feature of ellipticals is the increase of the 
central $<[Mg/Fe]>_{*}$ ratio with velocity dispersion 
(galactic mass, luminosity) ([Mg/Fe] vs. $\sigma_o$,
(Trager et al. 1993; Worthey et al. 1992;
Matteucci 1994; Jorgensen 1999; Kuntschner et al.  2001) which indicates that
more massive objects evolve faster than less massive ones.

\item Lyman-break and SCUBA galaxies at $z \ge 3$, 
where the star formation rate is as high as 
$\sim 40-1000 M_{\odot}yr^{-1}$, can be the young ellipticals
(Dickinson 1998; Pettini et al. 2002; De Mello et al. 2002; 
Matteucci \& Pipino 2002).

\item The existence of old fully assembled massive 
spheroidals already at $1.6 \le z \le 1.9$
(Cimatti et al. 2004) also indicate an early formation of ellipticals, at 
least at $z> 2$.

\item 
Very recently the Hubble Space Telescope has provided evidence for the 
existence of old massive spheroids at very high redshift. In particular, 
Mobasher et al. (2005) reported  
evidence for a massive ($M=6 \cdot 10^{11}M_{\odot}$) post-starburst galaxy at
$z=6.5$. 
\end{itemize}

These observational facts, in particular the high-z old and massive early type 
galaxies are challenging most N-body and semi-analitical simulations published 
so far, where these galaxies are very rare objects. 
Differences between predictions and observations are as high as a factor 
of ten and increase with z (Somerville et al. 2004). In addition,
evidence for mass downsizing and ``top-down'' assembly of ellipticals
(Cimatti et al. 2006; Renzini 2006) arises from a new analysis 
of the rest-frame B-band COMBO-17 and DEEP2 luminosity functions and from a photometric analysis of 
galaxies at z=1 (Kodama et al. 2004).
Therefore, all of these findings are pointing to a formation of ellipticals at very high redshift.

On the other hand, the arguments favoring the formation of ellipticals at 
low redshift  can be summarized as:  
\begin{itemize}

\item The relative large values of the $H_{\beta}$ index measured in a 
sample of nearby ellipticals which could  indicate prolonged star formation 
activity up to 2 Gyr ago (Gonzalez 1993; 
Trager et al. 1998).

\item The blue cores found in some ellipticals in the Hubble Deep Field 
(Menanteau et al., 2001) 
which indicate continuous star formation.

\item  
The tight relations in the fundamental plane at low and higher redshift 
can be interpreted as due to a conspiracy 
of age and metallicity, namely to an age-metallicity anticorrelation: 
more metal rich galaxies are younger than less metal rich ones
(e.g. Worthey et al. 1995; Trager et al. 1998,2000; Ferreras et al. 1999).

\item The main argument in favor of the formation at low redshift was for 
years the apparent paucity of high luminosity ellipticals 
at $z \sim 1$ compared to now (e.g. Kauffmann et al. 1996; 
Zepf, 1997; Menanteau et al. 1999) However, more 
recently Yamada et al. (2005) found that 
60-85\% of the local early type galaxies are  already in place at z=1.

\end{itemize}

In the following we will show the predictions of different models for the 
formation and evolution of ellipticals and compare these predictions with 
the above described observations.

\section{Models based on galactic winds}

Monolithic models assume that ellipticals suffer a 
strong star formation
and quickly produce galactic winds when the energy from SNe injected 
into the ISM equates the potential energy of the gas. Star formation 
is assumed to halt after the development
of the galactic wind and the galaxy is assumed to evolve passively afterwards.
The original model of Larson (1974) suggested that galactic winds should occur 
later in more massive objects due to the assumption of a constant efficiency 
of star formation in ellipticals of different mass and to the increasing 
depth of the potential well in more massive ellipticals. Unfortunately, this 
prediction is at variance with the observation that the $<[Mg/Fe]>_{*}$ ratio 
increases with galactic mass which instead suggests a shorter period of star 
formation for larger galaxies. This was first suggested by Trager et 
al. (1993), Worthey et al. (1994) and Matteucci (1994, hereafter M94)) who also computed 
models for ellipticals by assuming a shorter period of star formation in big 
ellipticals.In order to obtain that, an 
increasing efficiency of star formation with galactic mass was assumed with 
the consequence of obtaining a galactic wind occuring earlier in the massive 
than in the small galaxies. She called this process ``inverse wind'' and 
showed that such a model was able to reproduce the increase 
of $<[Mg/Fe]>_{*}$ with galactic mass.   

More recently, PM04 presented  a revised monolithic model
which allows for the formation of ellipticals by a fast merger of gas lumps 
at high redsfhit. The model is multizone and predicts that each elliptical 
forms ``outside-in'' (star formation stops in the outer before the 
inner regions qwing to a galactic wind). In other words, the galactic wind 
develops outside-in. 
Following the original suggestion by M94 they assumed an 
increasing efficiency of star formation with the galactic mass.
They also suggested a shorter timescale $\tau$ for the gas assembly with 
increasing galactic mass.
In Figure 1 we show the predicted histories of star formation in the ``inverse 
wind scenario'' of M94 and PM04. As one can see, the most massive 
ellipticals show a shorter and more intense episode of star formation than the 
less massive ones.

\begin{figure}
\includegraphics[width=5in,height=3.0in]{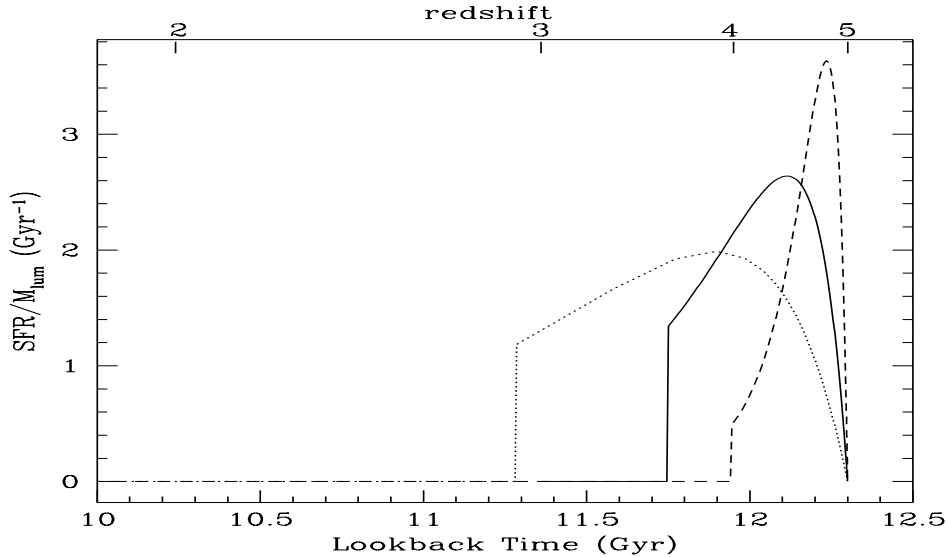}
\hfill
\caption{The predicted star formation histories (star formation rate per unit 
stellar mass) for galaxies of $10^{12}$ (dashed line), $10^{11}$ (continuos line) and $10^{10} M_{\odot}$ (dotted line). Such a behaviour is obtained by assuming that the efficiency of star formation is increasing with galactic mass whereas the timescale for the assembly of the gaseous lumps giving rise to the galaxies is a decreasing function of mass (downsizing both in star formation and mass assembly). In these models (PM04) the galactic wind occurs first in the more massive galaxies than in less massive ones.}\label{fig} 
\end{figure}

\begin{figure}
\includegraphics[width=5in,height=3.0in]{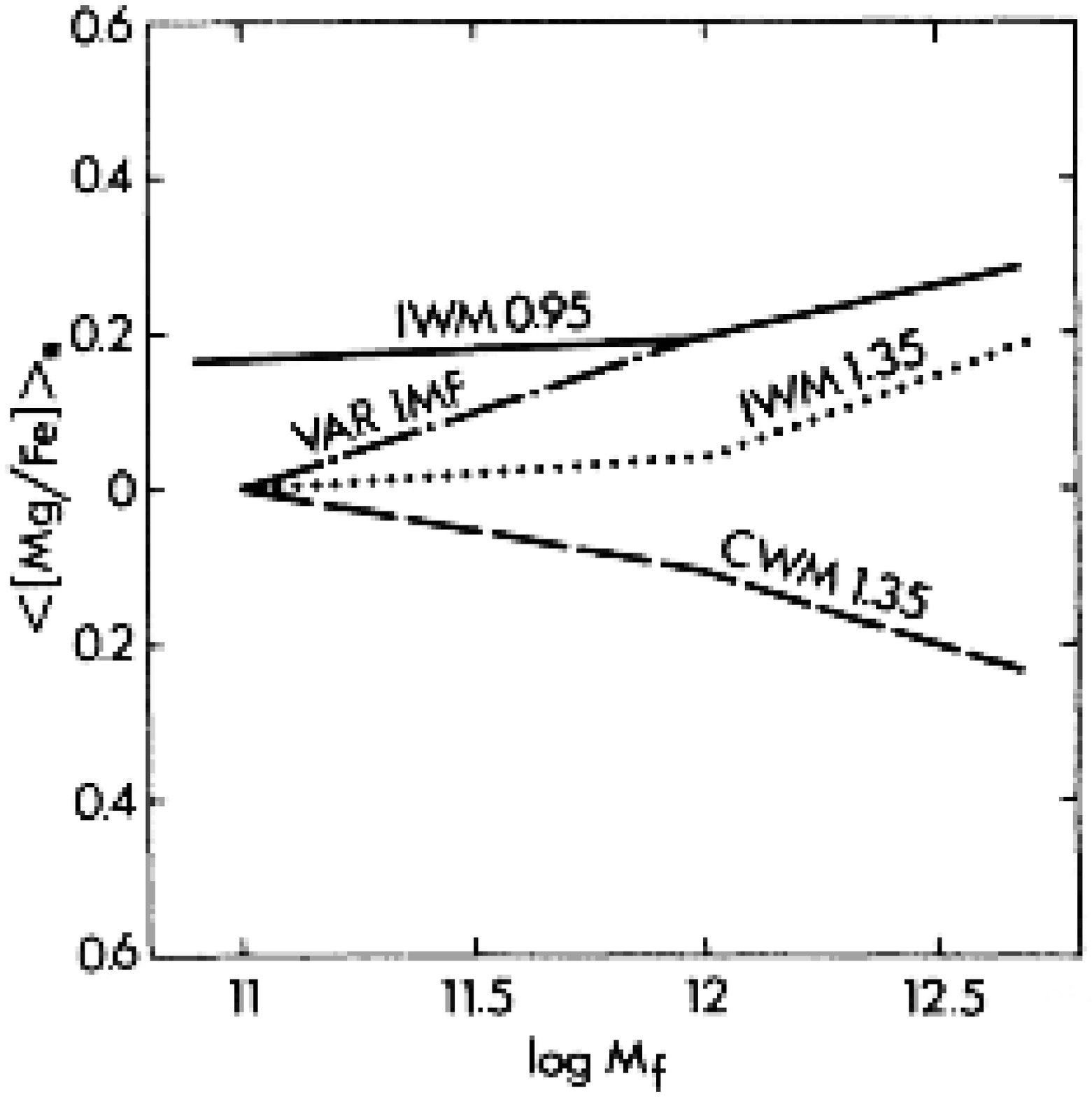}
\hfill
\caption{The predicted $<[Mg/Fe]>_{*}$ vs. log $M_f$ (final mass) for 
ellipticals, under several different assumptions by M94. The curves labelled IWM0.95 and IWM1.35 correspond to models with star formation histories similar to those shown in figure 1, the only difference being that the ellipticals are considered as closed-box systems until the occurrence of a galactic wind.The case IWM0.95 assumes for all galaxies an IMF with $x=0.95$, whereas the case IWM1.35 assumes a Salpeter (1955) IMF. The curve labeled CWM indicates classic wind models where the galactic wind occurs first in less massive than in more massive galaxies. Finally, the curve labeled VARIMF assumes a variable IMF as a function of the galactic mass (see text).}\label{fig} 
\end{figure}

In Figure 2 we show the predictions of Matteucci (1994) concerning the 
$<[Mg/Fe]>_{*}$ versus  the galactic mass (stellar) in the inverse 
wind scenario, the classical Larson's scenario and in the case of a variable 
initial mass function (IMF).
This last case assumes that more massive ellipticals should have a 
flatter IMF. However, this particular scenario requires a too flat IMF for 
massive ellipticals, at variance with observational properties 
(e.g. M/L ratio, colour-magnitude diagram).

\begin{figure}
\includegraphics[width=5in,height=3.0in]{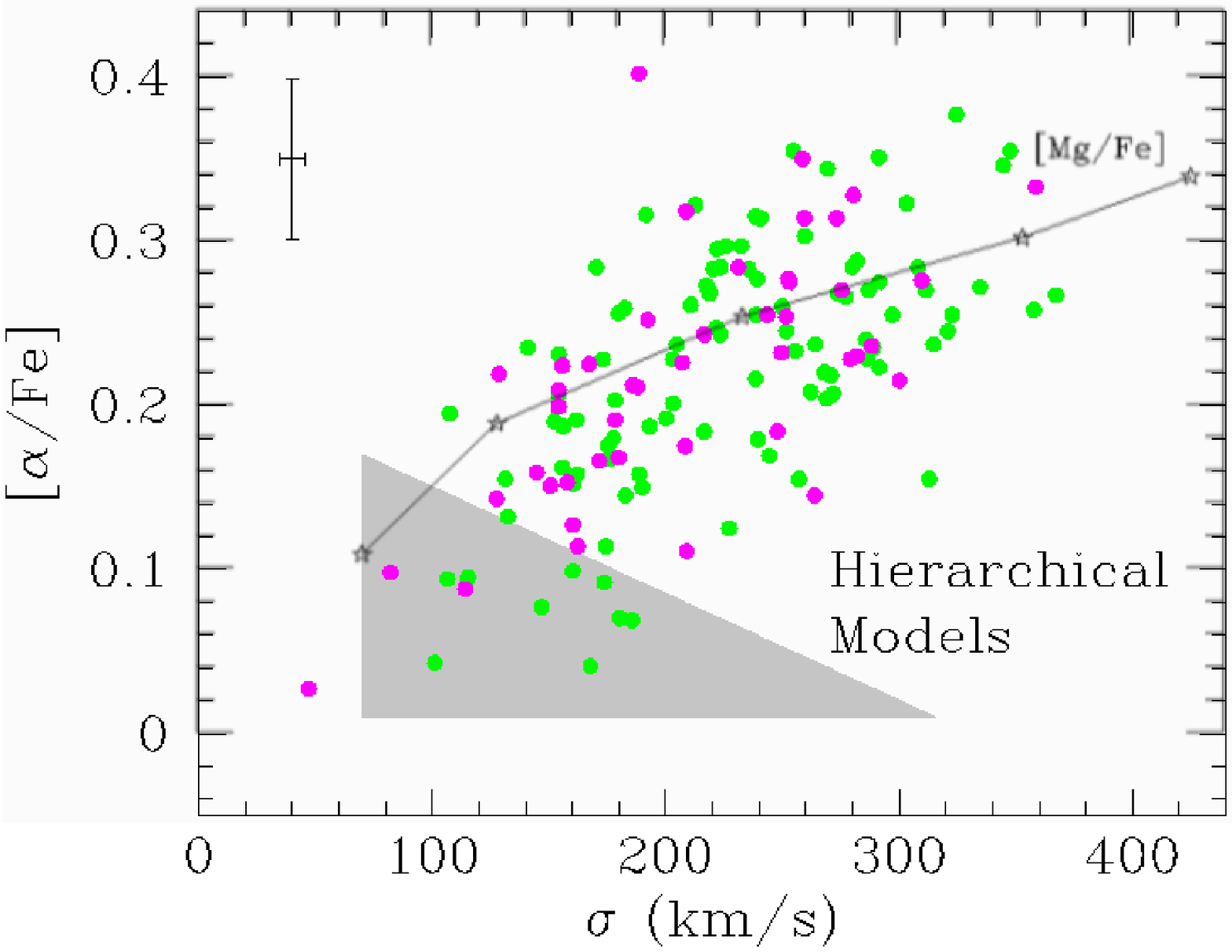}
\hfill
\caption{The [Mg/Fe] vs. $\sigma$ for ellipticals (continuos line) as 
predicted by PM04 compared with observations (dots) and with the 
predictions of 
hierarchical semi-analytical models (shaded area). Figure adapted from 
Thomas et al. (2002).}\label{fig} 
\end{figure}

PM04, recomputed the relation $<[Mg/Fe]>_{*}$ versus mass (velocity dispersion) and compared it with the data by Thomas et al. (2002), who
showed how hierarchical semi-analytical models cannot reproduce the observed $<[Mg/Fe]>_{*}$ vs. velocity dispersion trend, since in this scenario massive ellipticals have longer periods of star formation than smaller ones.
In Figure 3, 
we have plotted the predictions of PM04 (continuos line) compared with data and hierarchical clustering predictions. 

More recently, Thomas et al. (2005) presented a suggestion about the star formation histories in ellipticals, in the cases of high and low density environment (clusters and field) and suggested that the formation of ellipticals in the field might have started 2 Gyr after that of ellipticals in clusters. This suggestion is based on recent data relative to [$\alpha$/Fe] and ages in ellipticals and is shown in Figure 4. As one can see, their suggestion for the star formation in ellipticals in clusters is similar to that of PM04.

\begin{figure}
\includegraphics[width=5in,height=3.0in]{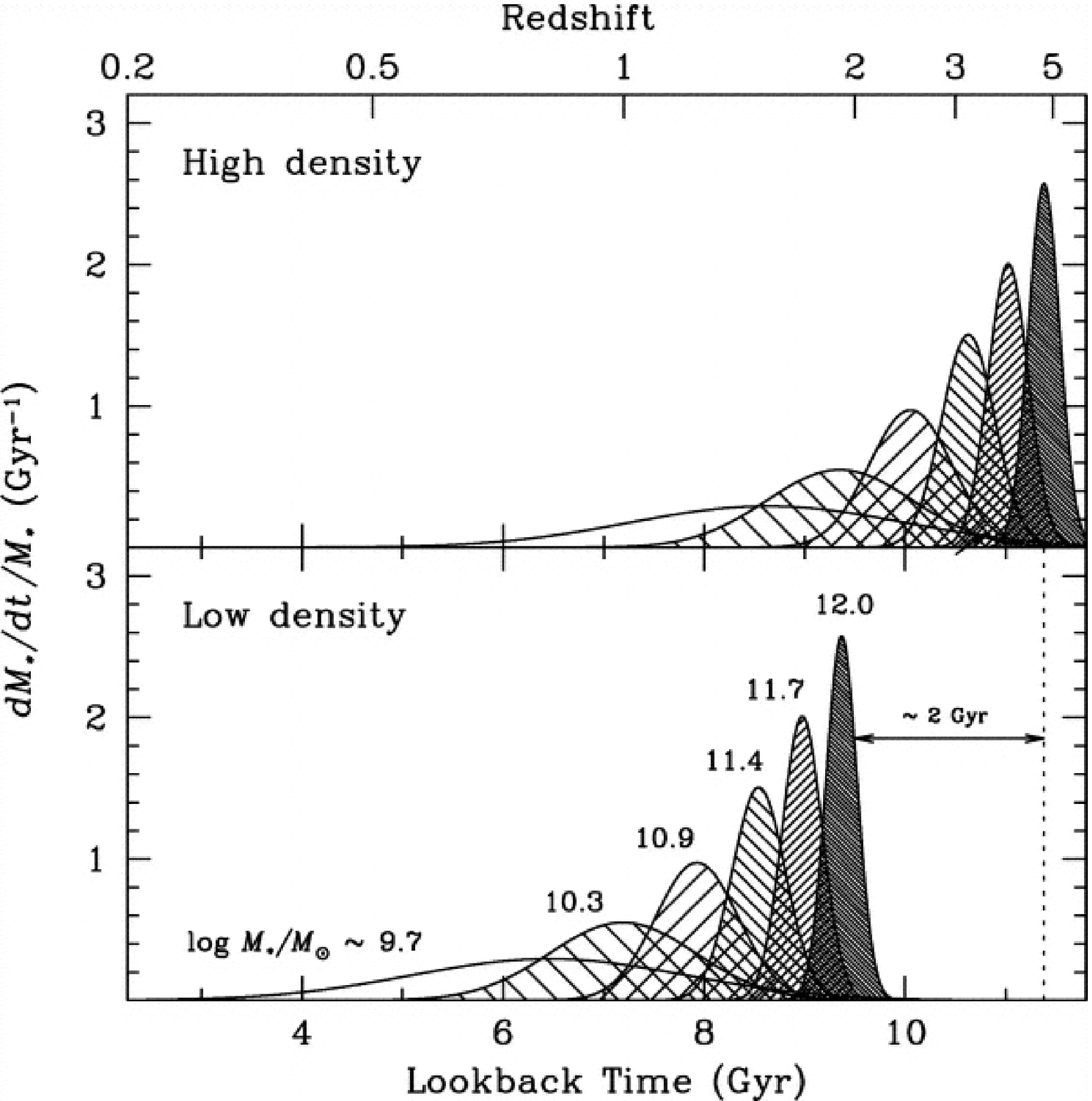}
\hfill
\caption{Thomas et al's view of the star formation history (star formation rate per unit mass of gas) in ellipticals of different masses and in different environments.Figure from Thomas et al. (2005).}\label{fig} 
\end{figure}

\section{Models with mergers}
In order to check the effect of possible gaseous mergers triggering star formation during the lifetime of elliptical galaxies on their chemical and photometric properties, Pipino \& Matteucci (2006) computed some cases with different merging epochs and different amounts of accreted matter.
In Figure 5 we show the predicted and observed $<[Mg/Fe]>_{V}$ vs.$M_V$ 
(visual magnitude) relation (the equivalent of the $<[Mg/Fe]>_{*}$ vs mass
relation) for ellipticals. The quantity $<[Mg/Fe]>_{V}$ represents the stellar 
[Mg/Fe] ratio averaged over $M_V$ instead of over the mass.

The figure contains the predictions of the best model of PM04 for galaxies of different mass, which lie well inside the area of existence of the observations, whereas the models with mergers tend to fall well outside the observed region unless the merger is unimportant. In particular, the agreement with observations worsens with the increasing merged mass and consequent star formation.

Thomas et al. (1999) studied a scenario where the formation of ellipticals 
occur by merging of two spirals like the Milky Way. They concluded that this scenario fails to reproduce the $\alpha$-enhanced abundance ratios in the metal rich stars of ellipticals, unless the IMF is flattened during the burst ignited by the merger.

\begin{figure}
\includegraphics[width=5in,height=3.0in]{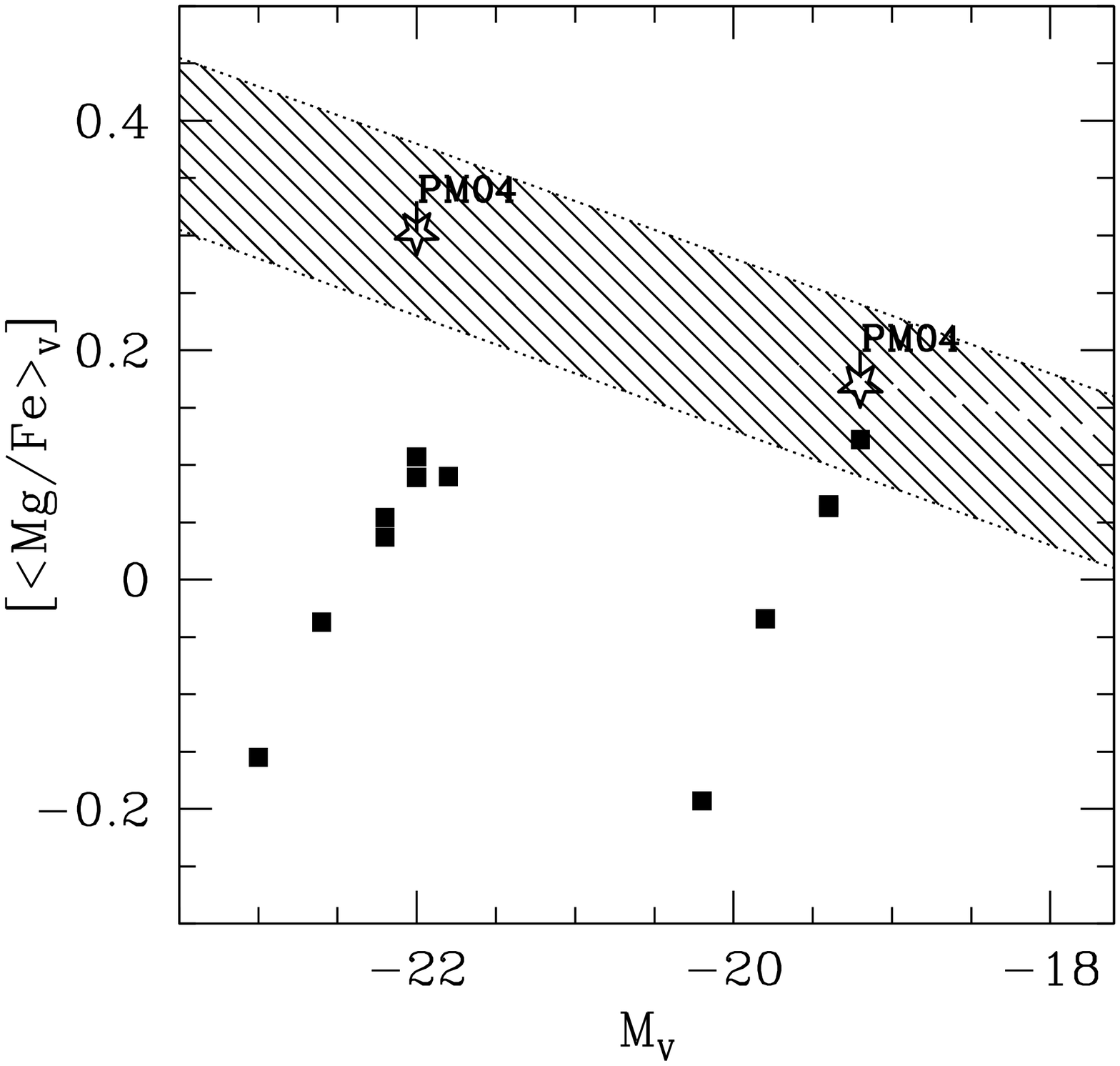}
\hfill
\caption{The observed $<[Mg/Fe]>_{V}$ vs. $M_{V}$ for ellipticals (shaded area)
compared with models without and with mergers. The big empty stars represent the best model of PM04 whereas the filled squares represent models with mergers. The squares lying more far from the shaded area represents models  where mergers with large amounts of gas and consequent star formation are allowed.}\label{fig}

\end{figure}

It is worth mentioning that recently, De Lucia et al. (2006) studied the star formation histories, ages and metallicities of ellipticals by means of the Millennium Simulation of the concordance $\Lambda$ CDM cosmology. They also suggested that more massive ellipticals should have shorter star formation timescales, but lower assembly (by dry mergers) redshift than less luminous systems.This is the first hierarchical paper admitting ``downsizing'' in the star formation process in 
ellipticals. However, the lower assembly redshift for the most massive system is still in contrast to what is concluded by Cimatti et al. (2006), who show that the downsizing trend should be extended also to the mass assembly, in the sense that the most massive ellipticals should have assembled before the less massive ones. This is in agreement with the model of PM04 which assumes an increasing timescale for the assembly of less massive ellipticals.

\section{The evolution of bulges}

The bulges of spiral galaxies are generally distinguishedd in true bulges, hosted by S0-Sb galaxies and ``pseudobulges'' hosted in later type galaxies
(see Renzini 2006 for references). Generally, the properties (luminosity, colors, line strenghts) of true bulges are very similar to elliptical galaxies.
In the following, we will refer only to true bulges and in particular to the Bulge of the Milky Way.
The Bulge of the Milky Way is, in fact, the best studied bulge and several scenarios for its formation have been put forward in past years.
As summarized by Wyse \& Gilmore (1992) the proposed scenarios are:
1) the Bulge formed by accretion of extant stellar systems which eventually 
settle in the center of the Galaxy.
2) The Bulge was formed by accumulation of gas at the center of the Galaxy and subsequent evolution with either fast or slow star formation.
3) The Bulge was formed by accumulation of 
metal enriched gas from the halo or thick disk or thin disk
in the Galaxy center.

The metallicity distribution of stars in the Bulge and the [$\alpha$/Fe] ratios
greatly help 
in selecting the most probable scenario for the Bulge 
formation. In Figure 6 we present the predictions by Matteucci (2001) of the 
[$\alpha$/Fe] ratios as functions of [Fe/H] in galaxies of different morphological type. In particular, for the  Bulge or an elliptical galaxy of the same mass, for the solar vicinity region and for an irregular magellanic galaxy (LMC and SMC).
The underlying assumption is that different objects undergo different  histories of star formation, being very fast in the spheroids (bulges and ellipticals), moderate in spiral disks and slow and perhaps gasping in irregular gas rich galaxies. The effect of different star formation histories is evident in Figure 6 where the predicted  [$\alpha$/Fe] ratios in the Bulge and ellipticals stay high and almost constant for a large interval of [Fe/H]. This is due to the fact that, since star formation is very intense, the Bulge reaches very soon a solar metallicity thanks only to the SNe II; then when SNe Ia start exploding and restoring Fe into the ISM, 
the change in the slope occurs at larger [Fe/H] than that in the solar vicinity.
In the extreme case of irregular galaxies the situation is opposite: here the star formation is slow and when the SNe Ia start exploding the gas is still very metal poor. 
This scheme is quite useful since it can be used to identify galaxies only by looking at their abundance ratios. 
A model for the Bulge behaving as shown in Figure 6 is able to reproduce also the observed metallicity distribution of Bulge stars (see Matteucci \& Brocato 1990; 
Matteucci et al. 1999).
The scenario suggested in these papers 
favors the formation of the Bulge by means of a short and strong starburst, in agreement with 
Elmegreen (1999) and Ferreras et al. (2003). A similar model, although updated with the inclusion of the development of a galactic wind and more recent stellar yields,  is presented by Ballero (this conference) who shows how a Bulge model with intense star formation 
(star formation efficiency $\sim 20 Gyr^{-1}$) and rapid assembly of gas (0.1 Gyr)
can best reproduce the most recent accurate data on abundance ratios and metallicity distribution.

\begin{figure}
\includegraphics[width=5in,height=3.0in]{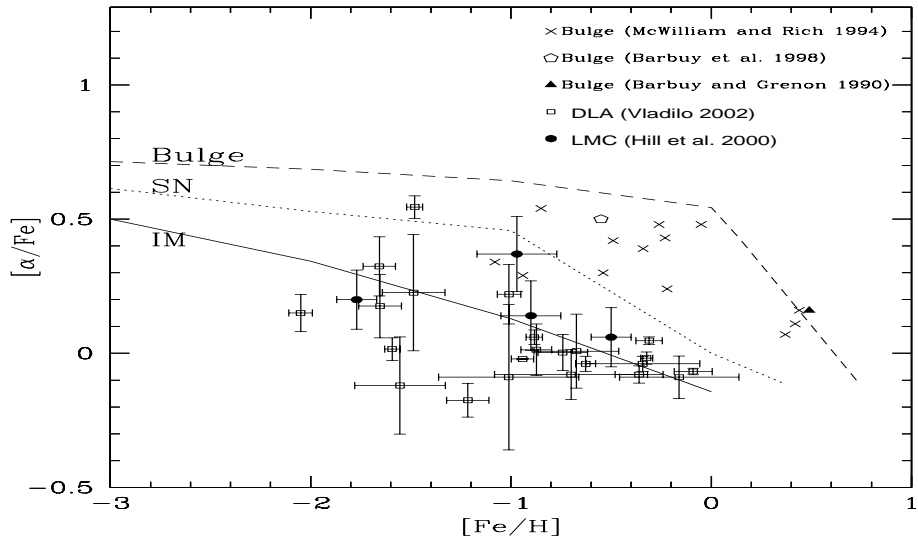}
\hfill
\caption{The predicted [$\alpha$/Fe] vs. [Fe/H] relations for the Bulge (upper curve), the solar vicinity (median curve) and irregular galaxies. Data for the Bulge are reported for comparison. Data for the LMC and Damped-Lyman $\alpha$ (DLA) Systems are also shown for comparison, indicating that DLA Systems are probably irregular galaxies. Figure from Matteucci (2001).}\label{fig}
\end{figure}

Previous attempts to model the galactic Bulge were also presented by Moll\'a et al. (2000) and by Samland et al. (1997).
Both these models, although different, assumed a more prolonged period of star formation than the models discussed above and this produces [$\alpha$/Fe] ratios behaving more akin to those in the solar vicinity.
Very recently, Zoccali et al. (2006) derived oxygen and iron abundances for  50 K giants in the galactic Bulge. The spectra are taken with UVES at the VLT and have a quite high  resolution (R=45,000). These data show a longer plateau for [O/Fe]  than in the solar vicinity, with a change in slope in the [O/Fe] vs. [Fe/H] relation occurring at [Fe/H] $ \sim -0.2$dex, in very good agreement with the predictions of Ballero et al. (2006, this conference). Also dynamical models by Immeli et al. (2004),  who simulate the formation of galaxies from clouds with different dissipation efficiencies, can explain the Bulge abundance ratios by means of a short starburst occurring when the dissipation efficiency is quite high.

\section{Conclusions}
We have compared the observational properties of ellipticals and bulges with 
the model predictions and reached the following conclusions:

\begin{itemize}

\item The existence of old and massive galaxies at high redshift ($z>3$)
argues in favor of the scenario where ellipticals form very rapidly and at high redshift.

\item The increasing $<[Mg/Fe]>_{*}$ with galactic mass as well as most 
of the properties of ellipticals 
(mass-metallicity relation, color-magnitude diagram)
can be well explained by a ``quasi-monolithic'' model where star 
formation stops 
first in more massive ellipticals because of galactic winds induced 
by supernovae. This implies a downsizing process both in the star formation and in the mass assembly of these galaxies. Models of co-evolution QSO-ellipticals also suggest also a shorter period of star formation in the most massive objects
(Granato et al. 2001).

\item
Classic hierarchical models for galaxy formation cannot reproduce the trend  
of [Mg/Fe] since they allow star formation 
to continue until recently in ellipticals. Only the recent model of De Lucia et al. (2006) combines a hierarchical scenario with star formation downsizing but it does not allow for a mass downsizing, as suggested by the recent work of Cimatti et al. (2006).

\item Major mergers occurring for elliptical galaxies at any redshift or late mergers with a variety of accreted 
masses, coupled with star formation, 
should be ruled out on the basis of the results of chemical and spectro-photometric models of ellipticals (Pipino \& Matteucci, 2006). Also mergers of spirals can be ruled out on the basis of chemical arguments (Thomas et al. 1999)

\item True bulges seem to be very old systems with properties similar to elliptical galaxies. 
The most recent galactic models and accurate abundance data
suggest that the Bulge of the Milky Way must have formed at very early times in a very short timescale ($< 0.5$ Gyr) and by means of a very intense burst of star formation with an efficiency of star formation.

\end{itemize}

\end{document}